\documentclass[twocolumn,showpacs,prd,amsmath,amssymb,nofootinbib,superscriptaddress]{revtex4}
\usepackage{graphicx,color}
\usepackage{amssymb,enumerate}
\newcommand\nn{\nonumber \\}

\newcommand{\lk}{\left(}
\newcommand{\rk}{\right)}
\newcommand{\ltk}{\left\{}
\newcommand{\rtk}{\right\}}
\newcommand{\ldk}{\left[}

\newcommand{\rdk}{\right]}
\newcommand\beq{ \begin{eqnarray} }
\newcommand\eeq{ \end{eqnarray} }

\begin{document}


\title{Color Magnetism in Non-Abelian Vortex Matter}

\author{Michikazu Kobayashi}
\email{michikaz@scphys.kyoto-u.ac.jp}
\affiliation{Department of Physics, Kyoto University, Oiwake-cho, Kitashirakawa, Sakyo-ku, Kyoto 606-8502, Japan}
\author{Eiji Nakano}
\email{e.nakano@kochi-u.ac.jp}
\affiliation{Department of Physics, Faculty of Science, Kochi University, Kochi 780-8520, Japan}
\author{Muneto Nitta}
\email{nitta@phys-h.keio.ac.jp}
\affiliation{Department of Physics, and Research and Education Center for Natural 
Sciences, Keio University, Hiyoshi 4-1-1, Yokohama, Kanagawa 223-8521, Japan}


\begin{abstract}
We propose color magnetism as a generalization of 
the ordinary Heisenberg (anti-)ferro magnets 
on a triangular lattice. 
Vortex matter consisting of 
an Abrikosov lattice of non-Abelian vortices 
with color magnetic fluxes 
shows a color ferro or anti-ferro magnetism,  
depending on the interaction among the vortex sites.
A prime example is a non-Abelian vortex lattice 
 in rotating dense quark matter, 
showing a color ferromagnetism. 
We show that the low-energy effective theory for the vortex lattice system 
in the color ferromagnetic phase 
is described by 
a $3+1$ dimensional ${\mathbb C}P^{N-1}$ nonlinear sigma model with spatially anisotropic couplings.  
We identify gapless excitations  
independent from Tkachenko modes
as color magnons, that is, 
Nambu-Goldstone modes 
propagating in the vortex lattice with an anisotropic 
linear dispersion relation $\omega_p^2 = c_{xy}^2(p_x^2+p_y^2)+c_z^2p_z^2$. 
We calculate the transition temperature between 
the ordered and disordered phases, 
and apply it to dense quark matter.
We also identify the order parameter spaces for color anti-ferromagnets.
\end{abstract}


\maketitle

\section{Introduction} \ 
Magnetism is one of important subjects in 
condensed matter physics 
because of fundamental question of what the origin of 
magnets is and 
applications to the modern technology. 
The magnetism should be explained from 
the electromagnetic interaction. 
On the other hand, 
there are other fundamental forces in nature, 
such as the strong interaction 
acting on quarks and gluons 
with color degrees of freedom. 
Here, we report a new state of matter which we call 
``color magnetism," 
that  may appear 
in non-Abelian vortex lattices 
when high density quark matter is rotating. 

Quark matter at finite temperature and/or 
density is likely to have a rich phase structure 
due to a variety of internal symmetries \cite{Fukushima:2010bq}.
At very high baryon number densities and low temperatures,
quark matter is expected to exhibit a color superconductivity 
where condensation of quark Cooper pairs breaks color symmetry 
as well as flavor and the baryon number symmetries. 
For three colors and three massless flavors, 
the order parameter of the quark pair condensation is given by 
\cite{Alford:1998mk,Alford:2007xm}
\beq
\Phi_{a \alpha}=\epsilon_{abc}\epsilon_{\alpha \beta \gamma}
\langle \psi^T_{b \beta}C\gamma_5  \psi_{c \gamma}\rangle=\Delta \delta_{a\alpha}, 
\eeq
where $ a,b,c$ ($\alpha,\beta,\gamma$) represent the flavor (color) indices, 
and $\Delta$ is the quark pairing gap. 
The symmetry breaking pattern,
${\rm SU}(3)_{\rm C}\times {\rm SU}(3)_{\rm L}\times {\rm SU}(3)_{\rm R} 
\times {\rm U}(1)_{\rm B}\rightarrow {\rm SU}(3)_{\rm C+L+R}$, 
shows that 
this phase retains only a simultaneous rotation of color and flavors as the residual symmetry, 
thus is called color-flavor locking (CFL) phase. 
We denote the suffix L+R by F in the following.

A possible candidate for system where the CFL phase takes place 
might be the core of very heavy compact stars, 
such as neuron stars. 
If the CFL phase develops in the rotating neutron star, 
vortices are created because of the superfluid nature  
associated with the $U(1)_{\rm B}$ breaking in the CFL phase. 
The most fundamental vortex in the CFL phase is a non-Abelian vortex 
\cite{Balachandran:2005ev,Eto:2009kg}, 
see Ref.~\cite{Eto:2013hoa} as a review.
The solution is of the form
\beq
\Phi=\Delta {\rm diag} \ltk e^{i\theta}f(r), g(r), g(r) \rtk
\eeq
where $(\theta, r)$ are the polar coordinates 
for a vortex extended along the $z$ direction, 
and $f$ and $g$ are profile functions with suitable boundary conditions.
Since the winding factor $e^{i\theta}$ is generated by 
both of the color $SU(3)_{\rm C}$ and $U(1)_{\rm B}$ rotations, 
the non-Abelian vortex carries a color-magnetic flux along the vortex core, 
and $1/3$ winding of the $U(1)_{\rm B}$ phase as well. 
Therefore, the non-Abelian vortex possesses nature of both superconducting and superfluid vortices, 
associated with breakings of the $SU(3)_{\rm C}$ color gauge symmetry 
and the baryon number $U(1)_{\rm B}$ symmetry, respectively. 

As the most remarkable property, the non-Abelian vortex has localized zero modes; 
the advent of the non-Abelian vortex further breaks the residual symmetry, 
${\rm SU}(3)_{\rm C+F}\rightarrow [{\rm SU}(2)\times U(1)]_{\rm C+F}$, 
and thus there appear the associated Nambu-Goldstone (NG) modes 
which are identified as ${\mathbb C}P^2$ orientation modes  
\cite{Nakano:2007dr,Eto:2009bh};
\beq
\frac{{\rm SU}(3)_{\rm C+F}}{[{\rm SU}(2)\times U(1)]_{\rm C+F}} 
\simeq {\mathbb C}P^2. 
\eeq
Since bulk side far from vortex core is approaching to the uniform CFL phase, 
the ${\mathbb C}P^2$ modes are normalizable, i.e., well localized along the vortex core. 
A different orientation of the ${\mathbb C}P^2$ modes corresponds to a different color of magnetic fluxes.
Since the ${\mathbb C}P^2$ modes are
transformed under $SU(3)_{\rm C+F}$ rotation as the fundamental representation, 
there are three eigenstates 
which we denote by a three-component complex variable $\phi$ as 
homogeneous coordinates of the ${\mathbb C}P^2$ space, 
where the $\phi$ satisfies a constraint $\phi^\dagger\phi=1$, 
and its overall phase factor is redundant by definition. 
Since the ${\mathbb C}P^2$ modes reside along the vortex line 
which we put along the $z$-axis, 
its low-energy effective theory on the vortex world sheet is 
obtained if the ${\mathbb C}P^2$ modes are promoted to $1+1$ dimensional fields $\phi(z,t)$ 
through the moduli space approximation \cite{Manton:1981mp}. 
It results in a nonlinear ${\mathbb C}P^2$ sigma model 
\cite{Eto:2009bh,Eto:2009tr},
\beq
{\mathcal L}_{{\mathbb C}P^2}=\sum_{\mu=0,3} K_\mu \ldk \partial^\mu \phi^\dagger \partial_\mu \phi 
+\lk \phi^\dagger \partial^\mu \phi \rk \lk \phi^\dagger \partial_\mu \phi \rk \rdk 
\label{eq:CPN}
\eeq
where $K_{0,3}$ are called the K\"ahler class 
given by the integration in terms of 
the profile functions $f$ and $g$.

However, 
the above ${\mathbb C}P^2$ model in 1+1 dimensions 
is gapped through quantum corrections 
\cite{Gorsky:2011hd,Eto:2011mk}, 
because of the Coleman-Mermin-Wagner theorem 
\cite{Coleman:1973ci,PhysRevLett.17.1133}
forbidding the presence of NG modes 
or a long range order in 1+1 dimensions. 

In this Letter, we show that this is only the case with a single vortex. 
When there is a vortex lattice consisting of 
a huge number of vortices, as expected in rotating quark matter 
such as in the core of neutron stars, 
there exist ${\mathbb C}P^2$ NG modes 
as gapless excitations in 3+1 dimensions 
whose dispersion relation is linear and 
anisotropic.
This is due to interactions between vortices 
depending on ${\mathbb C}P^2$ orientations \cite{Auzzi:2007wj}.
We discuss transition temperature between 
ordered and disordered phases and 
find that non-Abelian vortex matter form 
a color ferromagnet.

Apart from dense quark matter,  
non-Abelian vortices with 
${\mathbb C}P^{N-1}$ modes were found 
in supersymmetric $U(N)$ gauge theories with 
$N$ flavors \cite{Hanany:2003hp,Auzzi:2003fs} 
(see \cite{Eto:2006pg,Shifman:2007ce,2009supersymmetric} as a review).
The low-energy effective theory of a vortex 
is the ${\mathbb C}P^{N-1}$ model
which is in the same form with Eq.~(\ref{eq:CPN}) 
if we interpret $\phi$ as a complex $N$ vector.
Supersymmetric theories usually admit 
Bogomol'nyi-Prasado-Sommerfield (BPS) vortices 
among which no static interaction exist.
In this case, multiple vortices stably exist 
at arbitrary separations \cite{Eto:2005yh,Eto:2006cx}.
On the other hand, here we are interested in 
non-BPS vortices among which 
static force exist \cite{Auzzi:2007wj}. 
In the following, 
we consider general $N$.

\section{Non-Abelian vortex lattice} 

We discuss a color ferromagnetism of ${\mathbb C}P^{N-1}$ on a vortex lattice system. 
In rotating systems of CFL quark matter,
multiple non-Abelian vortices are created and may form a vortex lattice. 
In general, 
structure of the vortex lattice is determined by vortex-vortex interaction. 
In the present case of non-Abelian vortices, 
the interaction is mediated by massless $U(1)_{\rm B}$ phonons 
\cite{Nakano:2007dr,Nakano:2008dc}, 
and also by scalars and gluons \cite{Hirono:2010gq} which are massive, 
where the scalars are amplitude fluctuations of the gap function $\Phi$. 
If the vortex lattice is dilute, i.e., the inter vortex distance is large, 
the vortex-vortex interaction is dominated by the $U(1)_{\rm B}$ phonons 
which provide a long-range repulsion in the same manner of usual superfluid vortices. 
Thus the structure of the dilute vortex lattice must be the Abrikosov's 
triangular lattice \cite{Nakano:2007dr,Nakano:2008dc}.

We first define the system of our interest more precisely:
vortex lattice has the triangle structure of spacing $L$ in the $x$-$y$ plane, 
and these vortices are set parallel to each other along the $z$ direction. 
Once locations of vortices are fixed on the triangular lattice sites by the repulsive force, 
relative orientation of the ${\mathbb C}P^{N-1}$ modes between neighboring vortices are determined solely
by the interaction mediated by massive scalars and gluons. 
The ${\mathbb C}P^{N-1}$ dependent vortex-vortex interaction has already been derived 
in the previous study \cite{Eto:2013hoa}, 
and $V_{\rm int}^{\left<i,j\right>}$, the interaction energy per unit length in the $z$-direction, is of the form 
\beq
V_{\rm int}^{\left<i,j\right>} /\Delta^2 = \sum_A G_{\left<i,j\right>,A} 
\lk \phi_i^\dagger T_A \phi_i\rk \lk \phi_j^\dagger T_A \phi_j\rk, 
\label{cp2int1}
\eeq
where $\left<i,j\right>$ represents sites of two adjacent vortices, 
$G_{\left<i,j\right>,A}$ is the interaction strength between them, 
and $T_{A}$, $A=1,\cdots,N^2-1$, are generators of $SU(N)$. 
The above expression is the leading order result under assumption that
well separated two vortices are located parallel 
and their relative ${\mathbb C}P^{N-1}$ orientations are 
constant along vortex line. 
$G_{\left<i,j\right>,A}$ includes contributions from gluon-exchange and scalar-exchange potentials. 
The $U(1)_{\rm B}$ phonon does not feel the ${\mathbb C}P^{N-1}$ orientation, thus gives no contribution. 
In the context of quark matter, 
the gluon exchange part of the above ${\mathbb C}P^{N-1}$ dependent interaction was devised \cite{Eto:2013hoa} from 
the dual transformation which was used 
\cite{Hirono:2010gq}
to describe a topological interaction between non-Abelian vortex 
and quasiparticles such as gluons and phonon in the bulk space, 
while, at present, we do not have a such systematic derivation for coupling 
between the scalar fields and non-Abelian vortex. 
Nevertheless, a seminal work \cite{Auzzi:2007wj} 
on numerical simulation on the potential energy 
between two non-BPS non-Abelian vortices provides 
implication of an asymptotic form of full $G_{\left<i,j\right>,A}$ including both contributions. 
With inter vortex distance $r$ being defined and isotropy in the color space $A$, 
the asymptotic form of $G(r)=G_{\left<i,j\right>,A}$ would be given by \cite{Auzzi:2007wj}
\beq
G(r)\simeq -C_1 B_0(m_{\chi} r)+C_2 B_0(m_g r)
\label{af1}
\eeq
where $C_1$ and $C_2$ are positive numerical factors, 
and $B_0(r)$ the modified Bessel function, $B_0(mr)\sim \sqrt{\pi/2mr}e^{-mr}$ for $mr\gg 1$. 
In the above equation, the first term corresponds to massive scalar contributions 
and the second to massive gluon contributions, and 
it should be noted that the massive gluons and scalars give the opposite sign of interaction. 
For sufficiently large lattice spacing $L$, 
contribution of the lightest particle 
dominates the interaction.

The ${\mathbb C}P^{N-1}$ modes live in each vortex site, and their dynamics is 
described by the following effective Hamiltonian, 
\beq
&& H = \int dz\: \sum_{\left<i,j\right>,A} 
\Big[ -J_{xy} S_{i,A} S_{j,A} 
\nn
&& \hspace{1cm}
+ K_3 \{ |\partial_z \phi_i|^2 + (\phi_i^\dagger \partial_z \phi_i)^2 \} \Big] 
\label{eq:cpn-magnet} \\
&& \quad S_{i,A}:= \phi_i^\dagger T_A \phi_i,
\quad J_{xy} := \Delta^2 G(L)
\eeq
where $S_{i,A}$ are ``spin" variables 
and $J_{xy}$
is the interaction 
between neighboring vortices.

If $\phi$ is uniform in the $z$-direction, 
this Hamiltonian 
reduces to the Heisenberg model for $N=2$.
In this case, 
$J_{xy}>0$ and $J_{xy}<0$ correspond to ferromagnets
and  anti-ferromagnets, respectively. 
We use these terminologies for general $N$.

The model with $N \geq 3$ further provides
higher-dimensional nontrivial spin system as a new statistical model.
The case of $N=3$ corresponds to 
color magnetism in rotating dense quark matter discussed here.
First, the K\"{a}hler class $K_{0,3}$ is estimated as 
$K_0 \simeq 3K_3 \simeq 3\mu^2/\Delta^2$ at very high density 
\cite{Eto:2009tr}.  
The interaction $J_{xy}$ is determined through 
the masses of particles 
which vortices exchange.
At very high densities, 
the masses are given as 
$m_s=2m_\chi \sim 2\Delta , \; 
m_g \sim g \mu$,
where $m_{s,\chi}$ corresponds to the singlet (adjoint) part of scalar fields 
in the residual $SU(3)_{\rm C+F}$ symmetry, 
$\mu$ is the baryon number chemical potential, and
$g$ is coupling constant of strong interaction. 
At very high densities, i.e., very large $\mu$, $g$ and $\Delta$ are exponentially small 
with respect to $\mu$, therefore, one expects 
$m_g\gg m_s, m_\chi$.
Thus,  
\beq
G(L)\simeq -C_1 \sqrt{\frac{\pi}{2m_\chi L}}e^{-m_\chi L}. 
\eeq
Therefore, in the case of rotating CFL quark matter, 
because of the sign of the interaction above, 
the nearest neighbor interaction of ${\mathbb C}P^{N-1}$ modes 
is found to align their orientations (color fluxes), 
leading to a color ferromagnetism. 

An estimation of $J_{xy}$ is given as follows. 
Supposing the present system is inside of a neutron star of the radius $R$
with rotation period $P_{\rm rot}$, 
we can valuate a ballpark of the number of vortex as  
$N_v\simeq 1.9 \times 10^{19}\lk\frac{1{\rm ms}}{P_{{\rm rot}}}\rk 
\lk\frac{\mu/3}{300{\rm MeV}}\rk \lk\frac{R}{10{\rm km}}\rk$, from which 
the lattice spacing is estimated \cite{Hirono:2012ki} 
as $L\equiv \lk\frac{\pi R^2}{N_{v}}\rk^{1/2}
\simeq 4.0\times 10^{-6}{\rm m}
\lk\frac{P_{{\rm rot}}}{1{\rm ms}}\rk^{1/2}\lk\frac{300{\rm MeV}}{\mu/3}\rk^{1/2}$.
A set of values of $C_1\sim 10$, $L\simeq 4.0\times 10^{-6}$m, and 
the gap energy $\Delta\sim 100$ MeV for low temperatures  
~\cite{Alford:1998mk,Hessels:2006ze}
gives $\log\lk J_{xy}/\Delta^2\rk \sim -10^9$, 
which is quite small in comparison to the coupling in the $z$ direction. 
To have $J_{xy}/\Delta^2=O(1)$, 
we need to put the system very close to the critical point 
$(T_{\mathrm{c}}^{\mathrm{CFL}} - T)/T_{\mathrm{c}}^{\mathrm{CFL}} \sim 10^{-18}$, 
where we have used a mean-field result 
$\Delta \simeq 2.16 T_{\mathrm{c}}^{\mathrm{CFL}}
 \sqrt{1-T/T_{\mathrm{c}}^{\mathrm{CFL}}}$~\cite{Iida:2003cc}, 
or we have to set an asymptotically high density. 

Here let us point out that 
the smallness of the effective coupling $J_{xy}$ in the above estimation 
is based upon multiple approximations, 
i.e., the moduli space approximation and an asymptotic form of the interaction (\ref{af1}), 
both of which work against correlations among orientions in the $x$-$y$ plane. 
What we claim here is that the present study provides 
the qualitative argument on how orientational modes make order of color magnetism, 
which is common feature for general non-BPS vortex lattice systems \cite{Auzzi:2007wj}. 
So actual correlations amog orientations in the $x$-$y$ plane are more measurable, 
at least, for higher vortex densities. 
%

\section{Color magnons in color ferromagnets}
Low-energy excitations in the ordered phase have an important role in 
determination of, e.g., thermodynamic and transport properties of the system. 
The present system exhibits a color ferromagnetism,
thus we expect that there appear massless excitations like magnons in the ordinary ferromagnetism, 
which correspond to fluctuations of the ${\mathbb C}P^{N-1}$ modes 
around an orientation spontaneously set in color ferromagnetic phase. 
To investigate dynamics of these excitations, 
we take a long wavelength limit of the interaction between neighboring 
${\mathbb C}P^{N-1}$ modes, Eq.~(\ref{cp2int1}), 
which can be described by a continuum limit of the interaction. 
By use of Fierz transformation, 
$\sum_{A} \lk T_A\rk^a_b \lk T_A\rk^c_d$
$=$$\delta^a_d\delta^c_b/2-\delta^a_b\delta^c_d/2N$ 
($a,b,c,d=1,\cdots,N$),  
we can take a continuum limit of each bond of the triangle lattice in Eq.~(\ref{eq:cpn-magnet}), 
up to irrelevant constants, 
\beq
&&\sum_{A}
(\phi_{i}^\dagger T_A \phi_i)
\ (\phi_{j}^\dagger T_A \phi_j) \nn
\rightarrow &&
-
\frac{L^2}{2}\ldk |\nabla\phi|^2+\lk\phi^\dagger\nabla\phi\rk^2
+{\mathcal O}\lk \nabla^4\rk   \label{eq:CPN-lattice}
\rdk,  
\eeq
for a pair of the nearest neighbors $i,j$.
Then, we sum all bonds to obtain
\beq
&&
-\sum_{\left<i,j\right>,A}
J_{xy}\lk \phi_i^\dagger T_A \phi_i\rk \lk \phi_j^\dagger T_A \phi_j\rk \nn
&\rightarrow& 
\tilde{K}_{xy} \int {\rm d}x{\rm d}y \sum_{i=x,y}
\ldk |\partial_i \phi|^2
+\lk \phi^\dagger \partial_i \phi \rk^2 \rdk, \quad
\eeq
with a coupling constant being defined by 
$\tilde{K}_{xy}=J_{xy}(3L^2/2)(2/L^2\sqrt{3})(1/2)$$=$$\sqrt{3}J_{xy}/2$. 

Incorporating the effective theory in the $z$ and temporal directions 
with redefined couplings $\tilde{K}_{0,3}$$=$$2K_{0,3}/L^2\sqrt{3}$, 
where $2/L^2\sqrt{3}$ is the vortex density in the $x$-$y$ plane, 
we obtain a low-energy effective theory for 
the magnon modes in 3+1 dimension,   
$L_{\rm eff}= \int {\rm d}x{\rm d}y{\rm d}z{\mathcal L}_{\rm eff}$ with 
\beq
{\mathcal L}_{\rm eff} =
\sum_{\mu=0}^3 \tilde{K}_\mu \ldk \partial^\mu \phi^\dagger \partial_\mu \phi 
+\lk \phi^\dagger \partial^\mu \phi \rk \lk \phi^\dagger \partial_\mu \phi \rk \rdk 
\label{eq:spin-Lagrangian}
\eeq
which is a ${\mathbb C}P^{N-1}$ model 
with spatially anisotropic couplings; 
$\tilde{K}_1=\tilde{K}_2=\tilde{K}_{xy}\neq \tilde{K}_{3}$.

Now let us study the low-energy excitations of this model. 
To this end, we can set an orientation of the ground state as $\phi_0=\lk 1,0,\cdots,0 \rk$ 
without loss of generality. 
The color magnon corresponds to fluctuations around the orientation, 
$\phi=U(\delta_a)\phi_0=\phi_0+\tilde{\phi}+O(\tilde{\phi}^2)$, 
where $U(\delta_a)\in SU(N)_{\rm C+F}$ and 
$\tilde{\phi}:=i X_a \delta_a \phi_0=\lk i\delta_0, -\delta_1+i \delta_2,
\cdots, -\delta_{2N-3}+i \delta_{2(N-1)} \rk^T$.
Parameters $\delta_a$ are assigned to broken generators $X_a$ of the dimension of $SU(N)/SU(N-1)$,  
and are identified as the color magnons. 
The fluctuation $\delta_0$ is eliminated by overall $U(1)$ operation, thus will be redundant. 
Plugging these expression into the effective Lagrangian, 
we have dynamics of the color magnons as 
\beq
L_{\rm eff}&=&
\sum_{\mu=0}^3 \tilde{K}_\mu {\rm Tr} \ldk (1-\phi_0 \phi_0^\dagger)
\partial^\mu \tilde{\phi} \partial_\mu \tilde{\phi}^\dagger \rdk +O(\tilde{\phi}^3),  \nn
&=&
\sum_{\mu=0}^3 
\sum_{a=1}^{2(N-1)} \tilde{K}_\mu \partial^\mu \delta_a \partial_\mu \delta_a +O(\delta_a^3). 
\eeq
Then, the dispersion relation can be obtained as 
\beq
\omega_p^2 = 
(\tilde{K}_{xy}(p_x^2+p_y^2)+\tilde{K}_zp_z^2)/\tilde{K}_0, 
\eeq
which shows that the color magnons propagate in the $x$, $y$ directions 
with velocity $c_{xy}^2=\tilde{K}_{xy}/\tilde{K}_0$ 
and the $z$-direction with $c_{z}^2=\tilde{K}_{3}/\tilde{K}_0$. 
In the case of $N=3$ for the CFL quark matter, 
$c_{z}^2=K_{3}/K_0\simeq1/3$, and 
$c_{xy}^2=\Delta^4G(L)L^2\sqrt{3}/2\mu^2 
\simeq a_2 t^{7/4} e^{-a_1\sqrt{t}}$ where $t=(T_{\mathrm{c}}^{\mathrm{CFL}}-T)/T_{\mathrm{c}}^{\mathrm{CFL}}$, 
which is obtained from a mean-field result at very high densities. 
This formula gives the maximal value of $c_{xy}$ with respect to temperature, 
which is achieved at $t=(7/2a_1)^2$. 
For typical values of parameters, 
$a_1 \sim 10^{9}$, and $a_2\sim 10^{12.5}$,  
$c_{xy}\sim 10^{-9} (=0.3 {\rm m/s})$ at most,
giving 9 hours oscillation for a neutron star with 
$R=10$km. 

The color magnons are massless excitations propagating 
with a linear dispersion (of type I) 
independently from the other NG modes, 
e.g., Kelvin/Tkachenko modes associated with the breaking of translational invariance, 
which have a quadratic dispersion (of type II).

For a single non-Abelian vortex system, where the effective theory 
is given by the 1+1 dimensional ${\mathbb C}P^{N-1}$ model, 
there is no ordered phase because of strong quantum fluctuation effects.
For the present vortex lattice system, however, an ordered phase can be expected
because the fluctuations are suppressed by propagating along the $xy$-directions
in the long-wavelength limit, and
the mean-field analysis becomes better in such a 3+1 dimensional system compared to lower dimensional systems.
Within the mean-field analysis, the critical temperature $T_{\mathrm{c}}^{\mathrm{order}}$, 
below which the vortex lattice system is ordered,
can be estimated as 
(see Appendix A for derivation)
\begin{align}
T_{\mathrm{c}}^{\mathrm{order}} \sim \frac{J_{xy}}{k_{\mathrm{max}}} + K_3 k_{\mathrm{max}},
\end{align}
where
the first and second terms correspond to the mean-field critical temperature along the $xy$ and $z$-directions,  respectively, 
and $k_{\mathrm{max}}$ is
the maximum wavenumber taken into account along the $z$-direction.
In the continuum limit $k_{\mathrm{max}} \to \infty$ for the $z$-direction, the transition temperature diverges, 
which implies 
the ordered phase and the existence of the corresponding color magnon excitations in the whole temperature region. 
However, the divergent temperature is a mean-field artifact, 
and the disordered phase may be expected at the finite temperature 
within the effect beyond mean-field analysis.

\section{Color anti-ferromagnets}
Although the color magnetism with $J_{xy}>0$ 
is only the physical system with $N\geq 3$ known thus far, 
it may be worth 
to mention basic properties of color anti-ferromagnets with 
$J_{xy}<0$. 

We denote the order parameter space (OPS) for anti-ferromagnets 
by ${\cal M}^{N=2}_{\rm AF}$. We then find 
(see Appendix B for derivation)
\beq
&& 
{\cal M}^{N=2}_{\rm AF} \simeq SU(2)/{\mathbb Z}_2 \simeq SO(3)\\
&& 
{\cal M}^{N=3}_{\rm AF} \simeq SU(3)/U(1)^2  
\eeq
The OPS for $N=2$, $SO(3)$,  is well-known for Heisenberg anti-ferromagnets 
\cite{Kawamura:1984,Okubo:2010}
while the second is a so-called flag manifold of rank two.
The first and second homotopy groups of the OPS are 
important for the presence or absence of a Kosterlitz-Thouless (KT)
phase transition \cite{Kawamura:1984}. 
They are 
\beq
\pi_1({\cal M}^{N=2}_{\rm AF}) \simeq {\mathbb Z}_2 ,
&& \pi_2({\cal M}^{N=2}_{\rm AF}) \simeq 0 \\
\pi_1( {\cal M}^{N=3}_{\rm AF}) \simeq 0 ,
&& \pi_2( {\cal M}^{N=3}_{\rm AF}) \simeq {\mathbb Z} \oplus {\mathbb Z}  .
\eeq
The first homotopy groups show that the KT transition can 
happen only for $N=2$. 
The second homotopy groups imply 
the existence of two-dimensional Skyrmions
(lumps) extended as strings 
for $N = 3$.

\section{Summary and Discussion} 
As a generalization of the ordinary magnetism,
we have found color magnets
in lattices consisting of non-Abelian vortices 
with ${\mathbb C}P^{N-1}$ internal orientations.
While the case of $N=2$ reduces to Heisenberg (anti-)ferromagnets, 
rotating dense quark matter provides an example 
of color ferromagnets in the $N=3$ case. \footnote{
In response to an external magnetic field 
other vortex solution can be found, 
where gluons are condensed due to a chromomagnetic instability
\cite{Ferrer:2006ie,Ferrer:2007uw}.
Since the non-Abelian vortex we have discussed here 
generates a sponteneous magnetic field if the electromagnetism is introduced 
\cite{Vinci:2012mc}, 
investigation of their interplay is of interest as a future work.
}

For color ferromagnets, we have found color magnons as gapless NG modes propagating 
in 3+1 dimensions with the dispersion relation 
anisotropic in the direction along the rotating axis 
and the orthogonal plane, 
and have estimated the transition temperature 
in the mean-field approximation.
For color anti-ferromagnets, we have found the OPS and relevant homotopy groups. 

When we take into account the electromagnetic 
interactions $U(1)_{\rm EM}$, 
the ${\mathbb C}P^2$ modes of a 
non-Abelian vortex 
are electrically charged 
and the interaction between 
a vortex and the electromagnetic field  
is described by
a $U(1)_{\rm EM}$ gauged ${\mathbb C}P^2$ model 
\cite{Hirono:2012ki}. 
Hence, the ${\mathbb C}P^2$ modes 
affect the electric
conductivity in addition to the thermal conductivity.
There should be anisotropy on these conductivities, 
which may affect evolution of neutron stars 
such as the cooling process of the star, etc. 
As for electric conductivity, it was predicted that 
electro-magnetic waves along a vortex lattice decay 
and therefore the CFL phase acts as a polarizer 
\cite{Hirono:2012ki}.

Another effect of 
the electromagnetic 
interactions is a mixing between 
$U(1)_{\rm EM}$ and 
the $\lambda_8$ component of 
strong interaction.
This induces a tension difference 
among the ${\mathbb C}P^2$ degenerated vortices 
\cite{Vinci:2012mc}. This effect can be incorporated 
by an effective potential in 
the ${\mathbb C}P^2$ model.

A lattice formulation of the 
${\mathbb C}P^{N-1}$ model was proposed 
by using a $U(1)$ gauge field on link variables
\cite{Campostrini:1992ar,Campostrini:1992it}. 
Our model in Eq.~(\ref{eq:CPN-lattice}) provides 
a simpler lattice formulation without gauge fields.

\section*{Acknowledgements}

This work is supported in part by Grant-in-Aid for Scientific Research
(Grant No. 22740219 (M.K.),No. 24740166 (E.N.), and No. 25400268 (M.N.)) and the work of
M.N. is also supported in part by the `Topological Quantum Phenomena'' 
Grant-in-Aid for Scientific Research on Innovative Areas (No. 25103720)  
from the Ministry of Education, Culture, Sports, Science and Technology 
(MEXT) of Japan.


\appendix

\section{Transition Temperature within the Mean-Field Analysis}

We start from the system in the discrete layered triangular lattice:
\begin{align}
\begin{split}
H &= - \sum_{i \in (x,y,z),A} \bigg( J_{xy}^{\mathrm{dis}} \sum_{j \in NN_i(x,y)} S_{iA} S_{jA} \\
&\qquad\qquad\qquad + J_z^{\mathrm{dis}} \sum_{k \in NN_i(z)} S_{iA} S_{kA} \bigg), \label{eq:discrete-Hamiltonian}
\end{split}
\end{align}
where $S_{iA} \equiv \phi_i^\dagger T_A \phi_i$, $NN_i(x,y)$ is the nearest sites to $i$th site
for the triangular lattice in the $xy$-plane, and $NN_i(z)$ is the site in the nearest layer perpendicular to
the $z$-axis.
Dividing $S_{iA}$ into the mean field and its fluctuation, {\it i.e.}, $S_{iA} = \langle S_{A} \rangle + \delta S_{iA}$
and neglecting the second ordered fluctuation $\delta S_{iA} \delta S_{jA}$, we obtain
\begin{align}
\begin{split}
& H = \tilde{J} \sum_{A} \bigg( \frac{N_{xy} N_{z} \langle S_{A} \rangle^2}{2} - \sum_{i} \langle S_{A} \rangle S_{iA} \bigg) \\
& \tilde{J} \equiv (6 J_{xy}^{\mathrm{dis}} + 2 J_z^{\mathrm{dis}}), \\
\end{split}
\end{align}
where 6 and 2 in $\tilde{J}$ indicate the number of nearest sites
for the triangular lattice in the $xy$-plane and the layers perpendicular to the $z$-axis, and
$N_{xy}$ and $N_{z}$ are the total number of the triangular lattice sites and layers respectively.
The partition function $Z$ written as
\begin{align}
\begin{split}
& Z = e^{- \beta E_0} \int \prod_{i} \bigg(\prod_A D\psi_{iA}\bigg) \\
&\phantom{Z=\ } \times \delta\bigg(\sum_A |\psi_{iA}|^2 - 1\bigg) e^{\beta \tilde{J} \sum_A \langle S_{A} \rangle S_{iA}} \\
&\phantom{Z} \equiv e^{- \beta E_0} \prod_{i} Z_i,
\end{split} \\
& E_0 = E_0(\langle S_{A} \rangle) \equiv \frac{N_{xy} N_z \tilde{J}}{2} \sum_A \langle S_{A} \rangle^2,
\end{align}
can be calculated by diagonalizing $\sum_A \langle S_{A} \rangle S_{iA}$.
For ${\mathbb C}P^{N-1}$ case, $Z_i$ is formally written as $Z_i \equiv Z_N(B_1, \cdots, B_N)$
where, $B_i$ is the $i$th eigenvalue for $\sum_A \langle S_{A} \rangle S_{iA}$ satisfying $\sum_i B_i = 0$.
$Z_N$ satisfies the following recurrence relation:
\begin{align}
\begin{split}
&\phantom{=\ } \beta \tilde{J} Z_N(B_1, \cdots, B_N) \\
&= \frac{\pi}{B_{N-1} - B_{N}} \{Z_{N-1}(B_1, B_{N-2}, B_{N-1}) \\
&\phantom{\frac{\pi}{B_{N-1} - B_{N}}\{} - Z_{N-1}(B_1, B_{N-2}, B_{N})\},
\end{split} \\
& \beta \tilde{J} Z_2(B_1,B_2) = \frac{2 \pi^2(e^{B_1} - e^{B_2})}{B_1 - B_2}.
\end{align}
For $N = 3$, eigenvalues $B_1, B_2, B_3 = 1 - B_1 - B_2$ further satisfy $B_1^2 + B_2^2 + B_3^2 = 2 \beta \tilde{J} \sum_A \langle S_{A} \rangle^2$.
The free energy $F$ becomes
\begin{widetext}
\begin{align}
\begin{split}
\frac{F}{N_{xy} N_z \tilde{J}} &= \frac{B_1^2 + B_2^2 + B_1 B_2}{2} \\
&\phantom{=\ } - \frac{1}{\beta \tilde{J}} \Big[ \log\Big\{ (B_1 + 2 B_2) e^{\beta \tilde{J} B_1} + (B_1 - B_2) e^{- \beta \tilde{J} (B_1 + B_2)} - (2 B_1 + B_2) e^{\beta \tilde{J} B_2} \Big\} \\
&\phantom{= - \frac{1}{\beta \tilde{J}} [\ } - \log \{(B_1 - B_2) (B_1 + 2 B_2) (2 B_1 + B_2) \} + \log(2 \pi^3) - 2 \log(\beta \tilde{J}) \Big] \\
&\geqq \frac{3 B_1^2}{2} - \frac{1}{\beta \tilde{J}} \Big[ \log\Big\{ 1 + e^{3 \beta \tilde{J} B_1} (3 \beta \tilde{J} B_1 - 1)\Big\} \\
&\phantom{= \frac{3 B_1^2}{2} - \frac{1}{\beta \tilde{J}} \Big[ \ }
- 2 \beta \tilde{J} B_1 - 2 \log(3 B_1) + \log(2 \pi^3) - 2 \log(\beta \tilde{J}) \Big] \\
&= - \frac{3 \log(\pi)}{\beta \tilde{J}} + \frac{1}{4} (6 - \beta \tilde{J}) B_1^2 + \frac{\beta^2 \tilde{J}^2}{30} B_1^3 + \frac{\beta^3 \tilde{J}^3}{160} B_1^4 + O(B_1^5).
\end{split}
\end{align}
\end{widetext}
In the second line, the equality is hold for $B_2 = B_1$, $2 B_1$, or $B_1 / 2$.
With this free energy, the first ordered phase transition occurs at $T_{\mathrm{c}}^{\mathrm{order}} \simeq 53 \tilde{J} / 270
= 53 (3 J_{xy}^{\mathrm{dis}} + J_z^{\mathrm{dis}}) / 135$.

Next, we consider the continuum limit of the Hamiltonian \eqref{eq:discrete-Hamiltonian}.
Defining the distance between two layers as $\Delta_z = 1 / k_{\mathrm{max}}$ and
writing the continuous form as $\sum_{i \in z} \Delta_z \to \int dz$, we obtain the Hamiltonian \eqref{eq:discrete-Hamiltonian}
\begin{widetext}
\begin{align}
\begin{split}
H &= \frac{1}{\Delta_z} \sum_{i \in (x,y)} \int dz \Bigg[ - J_{xy}^{\mathrm{dis}} \sum_{j \in NN_i(x,y),A} S_{iA} S_{jA} 
+ 2 J_z^{\mathrm{dis}} \{ \Delta_z^2 |\partial_z \psi_i|^2 + \Delta_z^2 (\psi_i^\dagger \partial_z \psi_i)^2 
+ O(\Delta_z^4) \} \Bigg] \\
&\phantom{=\ } - 4 J^{\mathrm{dis}}_z N_{xy} N_z.
\end{split}
\end{align}
\end{widetext}
Compared to the original effective Hamiltonian \eqref{eq:cpn-magnet}, we get the coupling constants
\begin{align}
\begin{array}{c}
\displaystyle
J_{xy}^{\mathrm{dis}} = J_{xy} \Delta_z = \frac{J_{xy}}{k_{\mathrm{max}}}, \vspace{5pt} \\
\displaystyle
J_{z}^{\mathrm{dis}} = \frac{K_3}{2 \Delta_z} = \frac{K_3 k_{\mathrm{max}}}{2},
\end{array}
\end{align}
and the critical temperature
\begin{align}
T_{\mathrm{c}}^{\mathrm{order}} = \frac{159 J_{xy}}{135 k_{\mathrm{max}}} + \frac{53 K_3 k_{\mathrm{max}}}{270},
\end{align}
which diverges in the continuous limit $k_{\mathrm{max}} \to \infty$.

\section{OPS of color anti-ferromagnets on a triangular lattice}

Here we derive the OPS for color anti-ferromagnets on 
a triangular lattice by generalizing that of Heisenberg anti-ferromagnets.

First, we recall the case of $N=2$, 
Heisenberg anti-ferromagnets, 
in which the target space is ${\mathbb C}P^1 \simeq S^2$.
With picking up one point on a triangular lattice, 
we fix a point on $S^2$. 
We then look at two nearest points of a triangle.
The degrees of freedom to chose 
two points with an angle $2\pi/3$ with the first point of the $S^2$ 
is $S^1/{\mathbb Z}_2$. 
Then the OPS is $S^1/{\mathbb Z}_2$ fibered over $S^2$:
\beq
 {\cal M}^{N=2}_{\rm AF} \simeq S^1/{\mathbb Z}_2 \ltimes S^2 \simeq S^3/{\mathbb Z}_2 
\simeq SO(3). 
\eeq
The triangular lattice is filled by this set of triangle.

For higher $N\geq 3$, it is convenient to use 
homogeneous coordinates $\phi$ (a complex $N$-vector) of ${\mathbb C}P^{N-1}$.
We have a constraint $\phi^\dagger \phi=1$ and 
the overall phase is redundant. 

Let us consider the case of $N=3$,  
${\mathbb C}P^2 \simeq SU(3)/[SU(2)\times U(1)]$.
We take $\phi_1$ at one point of a triangular lattice 
to be $\phi_1^T = (1,0,0)$ without loss of generality.
Next, we take $\phi_2^T = (0,1,0)$ 
 $\phi_3^T = (0,0,1)$ without loss of generality.
Since the overall phase is redundant in each site, 
the symmetry $SU(3)$ is broken to $U(1)^2$. 
We then conclude 
\beq
{\cal M}^{N=3}_{\rm AF} \simeq SU(3)/U(1)^2 .
\eeq

\bibliography{dense-QCD-v2}

\begin{thebibliography}{34}
\expandafter\ifx\csname natexlab\endcsname\relax\def\natexlab#1{#1}\fi
\expandafter\ifx\csname bibnamefont\endcsname\relax
  \def\bibnamefont#1{#1}\fi
\expandafter\ifx\csname bibfnamefont\endcsname\relax
  \def\bibfnamefont#1{#1}\fi
\expandafter\ifx\csname citenamefont\endcsname\relax
  \def\citenamefont#1{#1}\fi
\expandafter\ifx\csname url\endcsname\relax
  \def\url#1{\texttt{#1}}\fi
\expandafter\ifx\csname urlprefix\endcsname\relax\def\urlprefix{URL }\fi
\providecommand{\bibinfo}[2]{#2}
\providecommand{\eprint}[2][]{\url{#2}}

\bibitem[{\citenamefont{Fukushima and Hatsuda}(2011)}]{Fukushima:2010bq}
\bibinfo{author}{\bibfnamefont{K.}~\bibnamefont{Fukushima}} \bibnamefont{and}
  \bibinfo{author}{\bibfnamefont{T.}~\bibnamefont{Hatsuda}},
  \bibinfo{journal}{Rept.Prog.Phys.} \textbf{\bibinfo{volume}{74}},
  \bibinfo{pages}{014001} (\bibinfo{year}{2011}), \eprint{1005.4814}.

\bibitem[{\citenamefont{Alford et~al.}(1999)\citenamefont{Alford, Rajagopal,
  and Wilczek}}]{Alford:1998mk}
\bibinfo{author}{\bibfnamefont{M.~G.} \bibnamefont{Alford}},
  \bibinfo{author}{\bibfnamefont{K.}~\bibnamefont{Rajagopal}},
  \bibnamefont{and} \bibinfo{author}{\bibfnamefont{F.}~\bibnamefont{Wilczek}},
  \bibinfo{journal}{Nucl.Phys.} \textbf{\bibinfo{volume}{B537}},
  \bibinfo{pages}{443} (\bibinfo{year}{1999}), \eprint{hep-ph/9804403}.

\bibitem[{\citenamefont{Alford et~al.}(2008)\citenamefont{Alford, Schmitt,
  Rajagopal, and Schafer}}]{Alford:2007xm}
\bibinfo{author}{\bibfnamefont{M.~G.} \bibnamefont{Alford}},
  \bibinfo{author}{\bibfnamefont{A.}~\bibnamefont{Schmitt}},
  \bibinfo{author}{\bibfnamefont{K.}~\bibnamefont{Rajagopal}},
  \bibnamefont{and} \bibinfo{author}{\bibfnamefont{T.}~\bibnamefont{Schafer}},
  \bibinfo{journal}{Rev.Mod.Phys.} \textbf{\bibinfo{volume}{80}},
  \bibinfo{pages}{1455} (\bibinfo{year}{2008}), \eprint{0709.4635}.

\bibitem[{\citenamefont{Balachandran et~al.}(2006)\citenamefont{Balachandran,
  Digal, and Matsuura}}]{Balachandran:2005ev}
\bibinfo{author}{\bibfnamefont{A.}~\bibnamefont{Balachandran}},
  \bibinfo{author}{\bibfnamefont{S.}~\bibnamefont{Digal}}, \bibnamefont{and}
  \bibinfo{author}{\bibfnamefont{T.}~\bibnamefont{Matsuura}},
  \bibinfo{journal}{Phys.Rev.} \textbf{\bibinfo{volume}{D73}},
  \bibinfo{pages}{074009} (\bibinfo{year}{2006}), \eprint{hep-ph/0509276}.

\bibitem[{\citenamefont{Eto and Nitta}(2009)}]{Eto:2009kg}
\bibinfo{author}{\bibfnamefont{M.}~\bibnamefont{Eto}} \bibnamefont{and}
  \bibinfo{author}{\bibfnamefont{M.}~\bibnamefont{Nitta}},
  \bibinfo{journal}{Phys.Rev.} \textbf{\bibinfo{volume}{D80}},
  \bibinfo{pages}{125007} (\bibinfo{year}{2009}), \eprint{0907.1278}.

\bibitem[{\citenamefont{Eto et~al.}(2013)\citenamefont{Eto, Hirono, Nitta, and
  Yasui}}]{Eto:2013hoa}
\bibinfo{author}{\bibfnamefont{M.}~\bibnamefont{Eto}},
  \bibinfo{author}{\bibfnamefont{Y.}~\bibnamefont{Hirono}},
  \bibinfo{author}{\bibfnamefont{M.}~\bibnamefont{Nitta}}, \bibnamefont{and}
  \bibinfo{author}{\bibfnamefont{S.}~\bibnamefont{Yasui}},
  \bibinfo{journal}{PTEP} \textbf{\bibinfo{volume}{2014}},
  \bibinfo{pages}{012D01} (\bibinfo{year}{2013}), \eprint{1308.1535}.

\bibitem[{\citenamefont{Nakano et~al.}(2008{\natexlab{a}})\citenamefont{Nakano,
  Nitta, and Matsuura}}]{Nakano:2007dr}
\bibinfo{author}{\bibfnamefont{E.}~\bibnamefont{Nakano}},
  \bibinfo{author}{\bibfnamefont{M.}~\bibnamefont{Nitta}}, \bibnamefont{and}
  \bibinfo{author}{\bibfnamefont{T.}~\bibnamefont{Matsuura}},
  \bibinfo{journal}{Phys.Rev.} \textbf{\bibinfo{volume}{D78}},
  \bibinfo{pages}{045002} (\bibinfo{year}{2008}{\natexlab{a}}),
  \eprint{0708.4096}.

\bibitem[{\citenamefont{Eto et~al.}(2009)\citenamefont{Eto, Nakano, and
  Nitta}}]{Eto:2009bh}
\bibinfo{author}{\bibfnamefont{M.}~\bibnamefont{Eto}},
  \bibinfo{author}{\bibfnamefont{E.}~\bibnamefont{Nakano}}, \bibnamefont{and}
  \bibinfo{author}{\bibfnamefont{M.}~\bibnamefont{Nitta}},
  \bibinfo{journal}{Phys.Rev.} \textbf{\bibinfo{volume}{D80}},
  \bibinfo{pages}{125011} (\bibinfo{year}{2009}), \eprint{0908.4470}.

\bibitem[{\citenamefont{Manton}(1982)}]{Manton:1981mp}
\bibinfo{author}{\bibfnamefont{N.}~\bibnamefont{Manton}},
  \bibinfo{journal}{Phys.Lett.} \textbf{\bibinfo{volume}{B110}},
  \bibinfo{pages}{54} (\bibinfo{year}{1982}).

\bibitem[{\citenamefont{Eto et~al.}(2010)\citenamefont{Eto, Nitta, and
  Yamamoto}}]{Eto:2009tr}
\bibinfo{author}{\bibfnamefont{M.}~\bibnamefont{Eto}},
  \bibinfo{author}{\bibfnamefont{M.}~\bibnamefont{Nitta}}, \bibnamefont{and}
  \bibinfo{author}{\bibfnamefont{N.}~\bibnamefont{Yamamoto}},
  \bibinfo{journal}{Phys.Rev.Lett.} \textbf{\bibinfo{volume}{104}},
  \bibinfo{pages}{161601} (\bibinfo{year}{2010}), \eprint{0912.1352}.

\bibitem[{\citenamefont{Gorsky et~al.}(2011)\citenamefont{Gorsky, Shifman, and
  Yung}}]{Gorsky:2011hd}
\bibinfo{author}{\bibfnamefont{A.}~\bibnamefont{Gorsky}},
  \bibinfo{author}{\bibfnamefont{M.}~\bibnamefont{Shifman}}, \bibnamefont{and}
  \bibinfo{author}{\bibfnamefont{A.}~\bibnamefont{Yung}},
  \bibinfo{journal}{Phys.Rev.} \textbf{\bibinfo{volume}{D83}},
  \bibinfo{pages}{085027} (\bibinfo{year}{2011}), \eprint{1101.1120}.

\bibitem[{\citenamefont{Eto et~al.}(2011)\citenamefont{Eto, Nitta, and
  Yamamoto}}]{Eto:2011mk}
\bibinfo{author}{\bibfnamefont{M.}~\bibnamefont{Eto}},
  \bibinfo{author}{\bibfnamefont{M.}~\bibnamefont{Nitta}}, \bibnamefont{and}
  \bibinfo{author}{\bibfnamefont{N.}~\bibnamefont{Yamamoto}},
  \bibinfo{journal}{Phys.Rev.} \textbf{\bibinfo{volume}{D83}},
  \bibinfo{pages}{085005} (\bibinfo{year}{2011}), \eprint{1101.2574}.

\bibitem[{\citenamefont{Coleman}(1973)}]{Coleman:1973ci}
\bibinfo{author}{\bibfnamefont{S.~R.} \bibnamefont{Coleman}},
  \bibinfo{journal}{Commun.Math.Phys.} \textbf{\bibinfo{volume}{31}},
  \bibinfo{pages}{259} (\bibinfo{year}{1973}).

\bibitem[{\citenamefont{Mermin and Wagner}(1966)}]{PhysRevLett.17.1133}
\bibinfo{author}{\bibfnamefont{N.~D.} \bibnamefont{Mermin}} \bibnamefont{and}
  \bibinfo{author}{\bibfnamefont{H.}~\bibnamefont{Wagner}},
  \bibinfo{journal}{Phys. Rev. Lett.} \textbf{\bibinfo{volume}{17}},
  \bibinfo{pages}{1133} (\bibinfo{year}{1966}).

\bibitem[{\citenamefont{Auzzi et~al.}(2008)\citenamefont{Auzzi, Eto, and
  Vinci}}]{Auzzi:2007wj}
\bibinfo{author}{\bibfnamefont{R.}~\bibnamefont{Auzzi}},
  \bibinfo{author}{\bibfnamefont{M.}~\bibnamefont{Eto}}, \bibnamefont{and}
  \bibinfo{author}{\bibfnamefont{W.}~\bibnamefont{Vinci}},
  \bibinfo{journal}{JHEP} \textbf{\bibinfo{volume}{0802}}, \bibinfo{pages}{100}
  (\bibinfo{year}{2008}), \eprint{0711.0116}.

\bibitem[{\citenamefont{Hanany and Tong}(2003)}]{Hanany:2003hp}
\bibinfo{author}{\bibfnamefont{A.}~\bibnamefont{Hanany}} \bibnamefont{and}
  \bibinfo{author}{\bibfnamefont{D.}~\bibnamefont{Tong}},
  \bibinfo{journal}{JHEP} \textbf{\bibinfo{volume}{0307}}, \bibinfo{pages}{037}
  (\bibinfo{year}{2003}), \eprint{hep-th/0306150}.

\bibitem[{\citenamefont{Auzzi et~al.}(2003)\citenamefont{Auzzi, Bolognesi,
  Evslin, Konishi, and Yung}}]{Auzzi:2003fs}
\bibinfo{author}{\bibfnamefont{R.}~\bibnamefont{Auzzi}},
  \bibinfo{author}{\bibfnamefont{S.}~\bibnamefont{Bolognesi}},
  \bibinfo{author}{\bibfnamefont{J.}~\bibnamefont{Evslin}},
  \bibinfo{author}{\bibfnamefont{K.}~\bibnamefont{Konishi}}, \bibnamefont{and}
  \bibinfo{author}{\bibfnamefont{A.}~\bibnamefont{Yung}},
  \bibinfo{journal}{Nucl.Phys.} \textbf{\bibinfo{volume}{B673}},
  \bibinfo{pages}{187} (\bibinfo{year}{2003}), \eprint{hep-th/0307287}.

\bibitem[{\citenamefont{Eto et~al.}(2006{\natexlab{a}})\citenamefont{Eto,
  Isozumi, Nitta, Ohashi, and Sakai}}]{Eto:2006pg}
\bibinfo{author}{\bibfnamefont{M.}~\bibnamefont{Eto}},
  \bibinfo{author}{\bibfnamefont{Y.}~\bibnamefont{Isozumi}},
  \bibinfo{author}{\bibfnamefont{M.}~\bibnamefont{Nitta}},
  \bibinfo{author}{\bibfnamefont{K.}~\bibnamefont{Ohashi}}, \bibnamefont{and}
  \bibinfo{author}{\bibfnamefont{N.}~\bibnamefont{Sakai}},
  \bibinfo{journal}{J.Phys.} \textbf{\bibinfo{volume}{A39}},
  \bibinfo{pages}{R315} (\bibinfo{year}{2006}{\natexlab{a}}),
  \eprint{hep-th/0602170}.

\bibitem[{\citenamefont{Shifman and Yung}(2007)}]{Shifman:2007ce}
\bibinfo{author}{\bibfnamefont{M.}~\bibnamefont{Shifman}} \bibnamefont{and}
  \bibinfo{author}{\bibfnamefont{A.}~\bibnamefont{Yung}},
  \bibinfo{journal}{Rev.Mod.Phys.} \textbf{\bibinfo{volume}{79}},
  \bibinfo{pages}{1139} (\bibinfo{year}{2007}), \eprint{hep-th/0703267}.

\bibitem[{\citenamefont{Shifman and Yung}(2009)}]{2009supersymmetric}
\bibinfo{author}{\bibfnamefont{M.}~\bibnamefont{Shifman}} \bibnamefont{and}
  \bibinfo{author}{\bibfnamefont{A.}~\bibnamefont{Yung}},
  \emph{\bibinfo{title}{Supersymmetric Solitons}}, Cambridge Monographs on
  Mathematical Physics (\bibinfo{publisher}{Cambridge University Press},
  \bibinfo{year}{2009}).

\bibitem[{\citenamefont{Eto et~al.}(2006{\natexlab{b}})\citenamefont{Eto,
  Isozumi, Nitta, Ohashi, and Sakai}}]{Eto:2005yh}
\bibinfo{author}{\bibfnamefont{M.}~\bibnamefont{Eto}},
  \bibinfo{author}{\bibfnamefont{Y.}~\bibnamefont{Isozumi}},
  \bibinfo{author}{\bibfnamefont{M.}~\bibnamefont{Nitta}},
  \bibinfo{author}{\bibfnamefont{K.}~\bibnamefont{Ohashi}}, \bibnamefont{and}
  \bibinfo{author}{\bibfnamefont{N.}~\bibnamefont{Sakai}},
  \bibinfo{journal}{Phys.Rev.Lett.} \textbf{\bibinfo{volume}{96}},
  \bibinfo{pages}{161601} (\bibinfo{year}{2006}{\natexlab{b}}),
  \eprint{hep-th/0511088}.

\bibitem[{\citenamefont{Eto et~al.}(2006{\natexlab{c}})\citenamefont{Eto,
  Konishi, Marmorini, Nitta, Ohashi et~al.}}]{Eto:2006cx}
\bibinfo{author}{\bibfnamefont{M.}~\bibnamefont{Eto}},
  \bibinfo{author}{\bibfnamefont{K.}~\bibnamefont{Konishi}},
  \bibinfo{author}{\bibfnamefont{G.}~\bibnamefont{Marmorini}},
  \bibinfo{author}{\bibfnamefont{M.}~\bibnamefont{Nitta}},
  \bibinfo{author}{\bibfnamefont{K.}~\bibnamefont{Ohashi}},
  \bibnamefont{et~al.}, \bibinfo{journal}{Phys.Rev.}
  \textbf{\bibinfo{volume}{D74}}, \bibinfo{pages}{065021}
  (\bibinfo{year}{2006}{\natexlab{c}}), \eprint{hep-th/0607070}.

\bibitem[{\citenamefont{Nakano et~al.}(2008{\natexlab{b}})\citenamefont{Nakano,
  Nitta, and Matsuura}}]{Nakano:2008dc}
\bibinfo{author}{\bibfnamefont{E.}~\bibnamefont{Nakano}},
  \bibinfo{author}{\bibfnamefont{M.}~\bibnamefont{Nitta}}, \bibnamefont{and}
  \bibinfo{author}{\bibfnamefont{T.}~\bibnamefont{Matsuura}},
  \bibinfo{journal}{Prog.Theor.Phys.Suppl.} \textbf{\bibinfo{volume}{174}},
  \bibinfo{pages}{254} (\bibinfo{year}{2008}{\natexlab{b}}),
  \eprint{0805.4539}.

\bibitem[{\citenamefont{Hirono et~al.}(2011)\citenamefont{Hirono, Kanazawa, and
  Nitta}}]{Hirono:2010gq}
\bibinfo{author}{\bibfnamefont{Y.}~\bibnamefont{Hirono}},
  \bibinfo{author}{\bibfnamefont{T.}~\bibnamefont{Kanazawa}}, \bibnamefont{and}
  \bibinfo{author}{\bibfnamefont{M.}~\bibnamefont{Nitta}},
  \bibinfo{journal}{Phys.Rev.} \textbf{\bibinfo{volume}{D83}},
  \bibinfo{pages}{085018} (\bibinfo{year}{2011}), \eprint{1012.6042}.

\bibitem[{\citenamefont{Hirono and Nitta}(2012)}]{Hirono:2012ki}
\bibinfo{author}{\bibfnamefont{Y.}~\bibnamefont{Hirono}} \bibnamefont{and}
  \bibinfo{author}{\bibfnamefont{M.}~\bibnamefont{Nitta}},
  \bibinfo{journal}{Phys.Rev.Lett.} \textbf{\bibinfo{volume}{109}},
  \bibinfo{pages}{062501} (\bibinfo{year}{2012}), \eprint{1203.5059}.

\bibitem[{\citenamefont{Hessels et~al.}(2006)\citenamefont{Hessels, Ransom,
  Stairs, Freire, Kaspi et~al.}}]{Hessels:2006ze}
\bibinfo{author}{\bibfnamefont{J.~W.} \bibnamefont{Hessels}},
  \bibinfo{author}{\bibfnamefont{S.~M.} \bibnamefont{Ransom}},
  \bibinfo{author}{\bibfnamefont{I.~H.} \bibnamefont{Stairs}},
  \bibinfo{author}{\bibfnamefont{P.~C.~C.} \bibnamefont{Freire}},
  \bibinfo{author}{\bibfnamefont{V.~M.} \bibnamefont{Kaspi}},
  \bibnamefont{et~al.}, \bibinfo{journal}{Science}
  \textbf{\bibinfo{volume}{311}}, \bibinfo{pages}{1901} (\bibinfo{year}{2006}),
  \eprint{astro-ph/0601337}.

\bibitem[{\citenamefont{Iida et~al.}(2004)\citenamefont{Iida, Matsuura,
  Tachibana, and Hatsuda}}]{Iida:2003cc}
\bibinfo{author}{\bibfnamefont{K.}~\bibnamefont{Iida}},
  \bibinfo{author}{\bibfnamefont{T.}~\bibnamefont{Matsuura}},
  \bibinfo{author}{\bibfnamefont{M.}~\bibnamefont{Tachibana}},
  \bibnamefont{and} \bibinfo{author}{\bibfnamefont{T.}~\bibnamefont{Hatsuda}},
  \bibinfo{journal}{Phys.Rev.Lett.} \textbf{\bibinfo{volume}{93}},
  \bibinfo{pages}{132001} (\bibinfo{year}{2004}), \eprint{hep-ph/0312363}.

\bibitem[{\citenamefont{Kawamura and Miyashita}(1984)}]{Kawamura:1984}
\bibinfo{author}{\bibfnamefont{H.}~\bibnamefont{Kawamura}} \bibnamefont{and}
  \bibinfo{author}{\bibfnamefont{S.}~\bibnamefont{Miyashita}},
  \bibinfo{journal}{J. Phys. Soc. Jpn.} \textbf{\bibinfo{volume}{53}},
  \bibinfo{pages}{4138} (\bibinfo{year}{1984}).

\bibitem[{\citenamefont{Okubo and Kawamura}(2010)}]{Okubo:2010}
\bibinfo{author}{\bibfnamefont{T.}~\bibnamefont{Okubo}} \bibnamefont{and}
  \bibinfo{author}{\bibfnamefont{H.}~\bibnamefont{Kawamura}},
  \bibinfo{journal}{J. Phys. Soc. Jpn.} \textbf{\bibinfo{volume}{79}},
  \bibinfo{pages}{085706} (\bibinfo{year}{2010}).

\bibitem[{\citenamefont{Ferrer and de~la Incera}(2006)}]{Ferrer:2006ie}
\bibinfo{author}{\bibfnamefont{E.~J.} \bibnamefont{Ferrer}} \bibnamefont{and}
  \bibinfo{author}{\bibfnamefont{V.}~\bibnamefont{de~la Incera}},
  \bibinfo{journal}{Phys.Rev.Lett.} \textbf{\bibinfo{volume}{97}},
  \bibinfo{pages}{122301} (\bibinfo{year}{2006}), \eprint{hep-ph/0604136}.

\bibitem[{\citenamefont{Ferrer and de~la Incera}(2007)}]{Ferrer:2007uw}
\bibinfo{author}{\bibfnamefont{E.~J.} \bibnamefont{Ferrer}} \bibnamefont{and}
  \bibinfo{author}{\bibfnamefont{V.}~\bibnamefont{de~la Incera}},
  \bibinfo{journal}{Phys.Rev.} \textbf{\bibinfo{volume}{D76}},
  \bibinfo{pages}{114012} (\bibinfo{year}{2007}), \eprint{0705.2403}.

\bibitem[{\citenamefont{Vinci et~al.}(2012)\citenamefont{Vinci, Cipriani, and
  Nitta}}]{Vinci:2012mc}
\bibinfo{author}{\bibfnamefont{W.}~\bibnamefont{Vinci}},
  \bibinfo{author}{\bibfnamefont{M.}~\bibnamefont{Cipriani}}, \bibnamefont{and}
  \bibinfo{author}{\bibfnamefont{M.}~\bibnamefont{Nitta}},
  \bibinfo{journal}{Phys.Rev.} \textbf{\bibinfo{volume}{D86}},
  \bibinfo{pages}{085018} (\bibinfo{year}{2012}), \eprint{1206.3535}.

\bibitem[{\citenamefont{Campostrini
  et~al.}(1992{\natexlab{a}})\citenamefont{Campostrini, Rossi, and
  Vicari}}]{Campostrini:1992ar}
\bibinfo{author}{\bibfnamefont{M.}~\bibnamefont{Campostrini}},
  \bibinfo{author}{\bibfnamefont{P.}~\bibnamefont{Rossi}}, \bibnamefont{and}
  \bibinfo{author}{\bibfnamefont{E.}~\bibnamefont{Vicari}},
  \bibinfo{journal}{Phys.Rev.} \textbf{\bibinfo{volume}{D46}},
  \bibinfo{pages}{2647} (\bibinfo{year}{1992}{\natexlab{a}}).

\bibitem[{\citenamefont{Campostrini
  et~al.}(1992{\natexlab{b}})\citenamefont{Campostrini, Rossi, and
  Vicari}}]{Campostrini:1992it}
\bibinfo{author}{\bibfnamefont{M.}~\bibnamefont{Campostrini}},
  \bibinfo{author}{\bibfnamefont{P.}~\bibnamefont{Rossi}}, \bibnamefont{and}
  \bibinfo{author}{\bibfnamefont{E.}~\bibnamefont{Vicari}},
  \bibinfo{journal}{Phys.Rev.} \textbf{\bibinfo{volume}{D46}},
  \bibinfo{pages}{4643} (\bibinfo{year}{1992}{\natexlab{b}}),
  \eprint{hep-lat/9207032}.

\end{thebibliography}

\end{document}